\documentclass[runningheads]{llncs}

\usepackage{tabularray}
\usepackage{color}
\usepackage{bbding}
\usepackage{subcaption,graphicx}
\usepackage{subcaption}
\usepackage{float}
\usepackage{eccv}

\usepackage{eccvabbrv}
\usepackage{wrapfig}
\usepackage{graphicx}
\usepackage{booktabs}
\definecolor{lightgray}{gray}{0.95}
\definecolor{color3}{gray}{0.95}
\definecolor{rouse}{rgb}{0.981,0.961,0.941}
\definecolor{rouse}{rgb}{0.981,0.961,0.941}
\usepackage[accsupp]{axessibility}  

\usepackage[pagebackref,breaklinks,colorlinks,citecolor=eccvblue]{hyperref}

\usepackage{orcidlink}
\usepackage{multirow}
\usepackage{booktabs}
\usepackage{colortbl}
\usepackage{makecell}                 

\begin{document}
\title{Exploiting Frequency Correlation for Hyperspectral Image Reconstruction} 

\author{Muge Yan\inst{1} \and
Lizhi Wang\inst{1}\and
Lin Zhu\inst{1}\\Hua Huang\inst{2}}
\authorrunning{Muge Yan et al.}

\institute{Beijing Institute of Technology \and
Beijing Normal University
}

\maketitle

\begin{abstract}
Deep priors have emerged as potent methods in hyperspectral image (HSI) reconstruction. While most methods emphasize space-domain learning using image space priors like non-local similarity, frequency-domain learning using image frequency priors remains neglected, limiting the reconstruction capability of networks. In this paper, we first propose a Hyperspectral Frequency Correlation (HFC) prior rooted in in-depth statistical frequency analyses of existent HSI datasets. Leveraging the HFC prior, we subsequently establish the frequency domain learning composed of a Spectral-wise self-Attention of Frequency (SAF) and a Spectral-spatial Interaction of Frequency (SIF) targeting low-frequency and high-frequency components, respectively. The outputs of SAF and SIF are adaptively merged by a learnable gating filter, thus achieving a thorough exploitation of image frequency priors. Integrating the frequency domain learning and the existing space domain learning, we finally develop the Correlation-driven Mixing Domains Transformer (CMDT) for HSI reconstruction. Extensive experiments highlight that our method surpasses various state-of-the-art (SOTA) methods in reconstruction quality and computational efficiency.
\end{abstract}

\section{Introduction}


\begin{wrapfigure}{t}{0.52\textwidth}
   \vspace{-16mm}
   \begin{center}
      \includegraphics[width=0.52\textwidth]{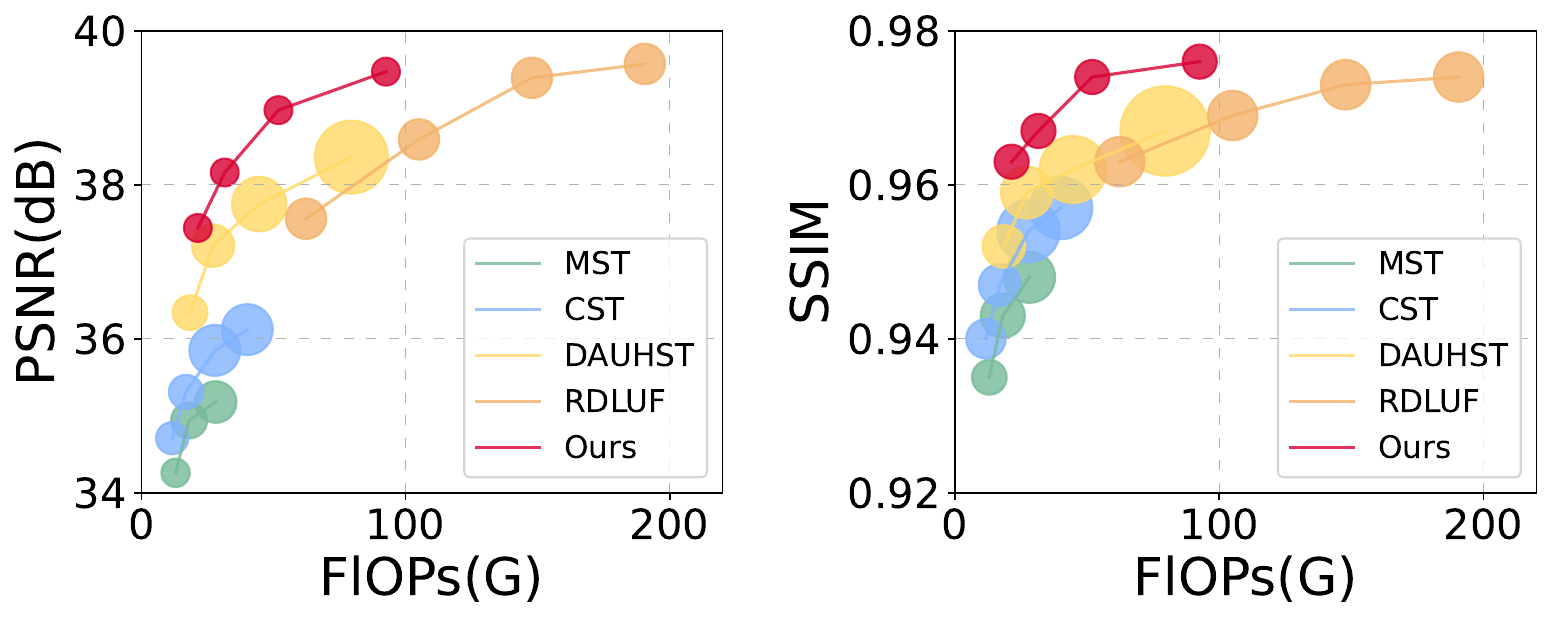}
   \end{center}
   \vspace{-12mm}
   \caption{\small PSNR-Params-FLOPs comparisons of our method and SOTA methods. The scale of circles matches with parameters (M).}
   \vspace{-10mm}
   \label{fig:psnr-flops}
\end{wrapfigure} 

Hyperspectral images (HSIs) encompass detailed spectral bands, thus widely used in image recognition~\cite{recognition1,recognition2}, tracking~\cite{nonlocal,track2}, and image classification~\cite{classification1,classification2,xu2023universal}. To capture HSIs, conventional spectrometers~\cite{james2007spectrograph,wolfe1997introduction} generally scan the scene along either the spatial dimension or the spectral dimension, requiring multiple exposures. Thus, these systems are unsuitable for measuring dynamic scenes. Based on the foundations of the compressive sensing theory~\cite{donoho2006compressed}, the Coded Aperture Snapshot Spectral Imaging (CASSI)~\cite{arce2013compressive,cassi,wang2015high,manakov2013reconfigurable} stands out as a promising solution. However, the bottleneck of CASSI lies in the limited reconstruction capacity of deriving the underlying 3D HSI from the 2D measurement.

For the inverse reconstruction problem, the HSI priors are essential to restrict the solution space. Traditional priors like sparsity~\cite{cassi} and low-rank~\cite{yuanlowrank} are hand-crafted and need manual parameter tweaking, thus lacking flexibility of processing different scenes. With the evolution of deep learning~\cite{wangcnn,wangdnn},
deep priors have emerged to be data-driven and flexible, thus contributing to a promoted reconstruction quality~\cite{wangcnn2,fuprior,dgsmp}.

\begin{wrapfigure}{t}{0.52\textwidth}
   \vspace{-16mm}
   \begin{center}
      \includegraphics[width=0.52\textwidth]{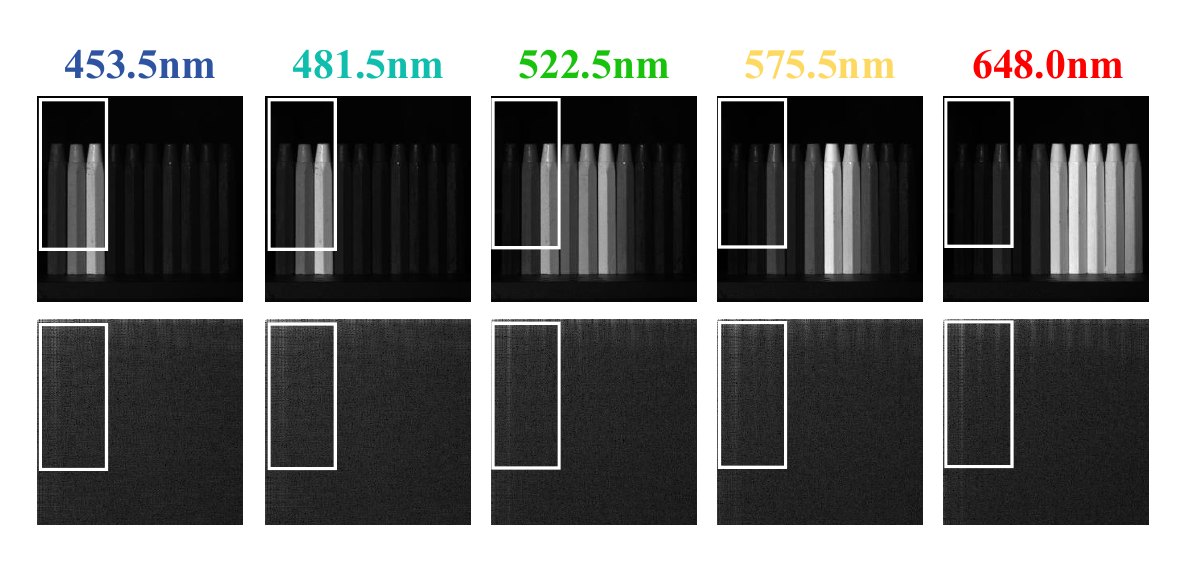}
   \end{center}
   \vspace{-12mm}
   \caption{\small The first row shows the gray images of \emph{Scene 9} in ascending spectral bands. The second row exhibits the spectrograms in ascending spectral bands corresponding to the first row.}
   \vspace{-10mm}
   \label{fig:roughview}
\end{wrapfigure} 

Nonetheless, the previous deep priors narrowly exploit space-domain information. However, existing space-domain-based models suffer from non-local issues with the limited receptive field of CNNs and inflexible fixed shuffle patterns of transformers, obstructing the learning of non-local similarity and frequency intricacies.
Furthermore, according to the various research on the inherent nature of neural networks~\cite{freq1,freq2,freq3}, the learning process of neural networks pays high attention to image frequency intricacies that represents the useful periodic patterns beyond space pixels, thus emphasizing the great promise to exploit the image frequency priors for HSI reconstruction. To harness the frequency information, HDNet~\cite{hdnet} first uses the focal frequency loss for network guidance. However, given that the frequency loss predominantly originates from low-frequency components, the inappropriate weighting manner of the focal frequency loss will inadvertently over-magnify low-frequency focus while over-diminishing high-frequency focus, thus resulting in suboptimal results with insufficient texture details.

In this paper, we first conduct a series of frequency analyses of HSIs and obtain a pivotal observation: a pronounced spectral-spatial correlation between frequency components,~\emph{i.e.}, tokens, are widely present in divergent HSIs and exhibits a descending trend from low-frequency to high-frequency token. Moreover, there exists the spectral/spatial correlation within one frequency token summarized from the neighborhood similarity. Thus, we develop the Hyperspectral Frequency Correlation (HFC) prior by consolidating the two-fold frequency correlations of HSIs into one overarching theme.

Leveraging the substantial HFC prior, we subsequently establish the detailed frequency domain learning to thoroughly exploit frequency correlation for HSI reconstruction. Concretely, the frequency domain learning is composed of two modules: the Spectral-spatial self-Attention of Frequency (SAF) and the Spectral/spatial Interaction of Frequency (SIF), which target the low-frequency tokens and the high-frequency tokens, respectively.
For the low-frequency tokens, given the dominant spectral-spatial correlation between frequency tokens, the SAF is creatively formulated based on the self-attention mechanisms to model the long-term dependencies between various spectral-spatial frequency tokens.
For the high-frequency tokens, as the high-frequency tokens exhibit subdued spectral-spatial correlation as illustrated in the HFC prior, it is inefficient to model the correlation between frequency tokens but of great significance to exploit the correlation within one frequency token. Thus, the SIF is concisely established with a spatial token interaction followed by a spectral token evolution, achieving mixed-dimensions exploration within one frequency token.
As the dominant correlation varies in different frequency components, a learnable gating filter is crafted to adaptively determine the weights of SAF and SIF and offer the optimal proportion of two blocks, ensuring thorough frequency exploitation.

By integrating the frequency domain learning and space domain learning, we finally formulate the Correlation-driven Mixing Domains Transformer (CMDT) as the heart of deep priors and subsequently obtain a deep frequency unfolding framework for HSI reconstruction. Extensive experiments on simulation data and real data validate the superiority of our method over various SOTA methods on both image quality and computational efficiency as shown in Fig.~\ref{fig:psnr-flops}.

Our contributions are summarized as follows:
\begin{itemize}
    \item We establish the useful HFC prior rooted in the statistical frequency analyses of the HSI datasets.
    
    \item We formulate the frequency domain learning comprising SAF and SIF to comprehensively exploit frequency correlation.
    
    \item We establish the CMDT comprising the frequency domain learning and space domain learning to efficiently explore the correlation of HSIs in dual domains.
    
\end{itemize}

\section{Related Work}
\label{sec:relatedWork}


\subsection{Image Priors for HSI}
Image priors are essential for high-quality HSI reconstruction. Traditional priors such as total variation~\cite{tv}, sparsity~\cite{cassi}, and low-rank~\cite{yuanlowrank} are hand-crafted. For instance, GAP-TV~\cite{gaptv} minimizes total variation to guarantee the first-order smoothness. DeSCI~\cite{desci} exploits repetitive patches under low-rank assumptions. However, these handcrafted priors require manual parameter tweaking, thus struggling with limited generalization. Inspired by the success of deep learning, deep priors have arisen to explore data-driven priors from large datasets. Wang et al.~\cite{wangprior} proposes a deep spectral-spatial network to explore the image spectral-spatial correlation in the space domain. DNU~\cite{dnu} introduces a non-local network to exploit the non-local similarity in the space domain. However, these priors solely focus on the image space priors, thus restricted in exploration of image frequency priors, resulting in a suboptimal reconstruction.

\subsection{Self-Attention Mechanism}
The self-attention mechanism~\cite{sam,yulunsam1,yulunsam2} has emerged as a powerful tool for capturing long-range interactions. Recently, self-attention mechanisms have showcased significant potential in image classification~\cite{classification1,classification2}, semantic segmentation~\cite{segmentation1,segmentation2}, image restoration~\cite{restoration1,restoration2,yulunsam2,yulunrestoration}, etc. For HSI reconstruction, $\lambda$-Net~\cite{lambdanet} first explores self-similarity via a self-attention mechanism. MST~\cite{mst} next calculates spectral self-attention map to discern image spectral correlation. DAUHST~\cite{dauhst} subsequently explores a shuffle strategy to calculate the image non-local correlation. However, the previous methods always leverage the self-attention mechanisms for the extraction of space-domain image correlation and struggle with striking a balance between low computational costs and modeling the comprehensive space-domain long-range dependencies, thus limiting the potential of the self-attention mechanism for HSI reconstruction.

\subsection{Spectral-spatial interaction}
Given the limited capacity of self-attention mechanisms, some works~\cite{interaction1, interaction2,interaction3} use the spectral-spatial interaction in the space domain to model the long-term dependency, which is mainly based on the common inductive biases like spatial non-local similarity and spectral dependency of space pixels. Concretely, SSIN~\cite{interaction1} designs a spectral-spatial attention module for interaction. ESSINet~\cite{interaction3} proposes an involution-2D operator to fuse the spectral and spatial features. In the frequency domain, ~\cite{Fourmer} attempts to migrate a similar architecture to the frequencies. However, the under-explored inductive biases of frequency make the frequency learning process lack reasonable interpretability, thus suppressing the future potential of frequency-domain learning.

\subsection{Learning in the Frequency Domain}
Frequency analysis~\cite{freqAna1,freqAna2} emphasizes signal frequency attributes. Numerous studies have probed the frequency traits of neural networks to explore network generalization. F-principle~\cite{freq1} elucidates that network learning inherently obeys an order from low to high frequencies, which is concordant with the human visual system~\cite{freq2} emphasizes that the generalization mainly originated from frequency learning, illustrating the importance of learning the frequency information of images. These insights have galvanized endeavors in frequency exploration. For instance, ~\cite{focalfreqloss} proposes a focal frequency loss for network guidance. DASR~\cite{dasr} and SFNet~\cite{selective} deploy filters to grasp frequency features from spatial images. ~\cite{freqrestoration} proposes the reasonable mathematical method of frequencies for HSI restoration. However, prevalent models~\cite{selective, hdnet, dasr, DFLFME} grapple with either elucidating the mechanics of frequency extraction from space-domain images or lacking learning adaptability of frequencies in HSI, underscoring the ongoing need for an interpretable and efficient frequency-domain network.

\section{The Proposed Method}
\vspace{-1em}
\label{sec:method}
In this section, we first introduce the Hyperspectral Frequency Correlation (HFC) prior. Based on the HFC, we next formulate the novel frequency domain learning block for thorough frequency exploitation. Integrating the space domain learning, we subsequently develop the Correlation-driven Mixing Domains Transformer (CMDT) for comprehensive image prior exploration. By plugging CMDT into an unfolding architecture, we finally form a deep frequency unfolding framework for HSI reconstruction.


\vspace{-1em}
\subsection{Hyperspectral Frequency Correlation Prior}
\vspace{-0.5em}
Among various techniques for transforming image signals from the space domain to the frequency domain, Discrete Cosine Transform (DCT) generates a frequency coefficient matrix comprising entirely real values, making it convenient for frequency analysis. Therefore, we conduct a series of analyses on the spectrogram after the 2D-DCT.

\noindent\textbf{Spectral/spatial Correlation within One Frequency Token.} Due to the unique arrangement of frequency in the spectrogram, the frequencies within a spatial neighborhood are of similar horizontal and vertical wavelengths, corresponding to the close positions in the DCT coefficient matrix. Thus, for the 2D-spectrogram of an HSI in one spectral band, a correlation emerges within one frequency token in a spatial neighborhood. Moreover, for a specific frequency component, a spectral correlation also exists within the frequency token in a spectral group. We term two correlations in both spatial neighborhood and spectral group spectral/spatial correlation within one frequency token.

\begin{figure}[tp]
   \vspace{-8mm}
   \begin{center}
   \begin{subfigure}{0.38\textwidth}
      \includegraphics[width=\textwidth]{figure/HFC.pdf}
      \caption{\small Correlation heatmaps}
      \label{fig:subfig1}
    \end{subfigure}
    \hspace{5mm}
    \begin{subfigure}{0.50\textwidth}
        \includegraphics[width=\linewidth]{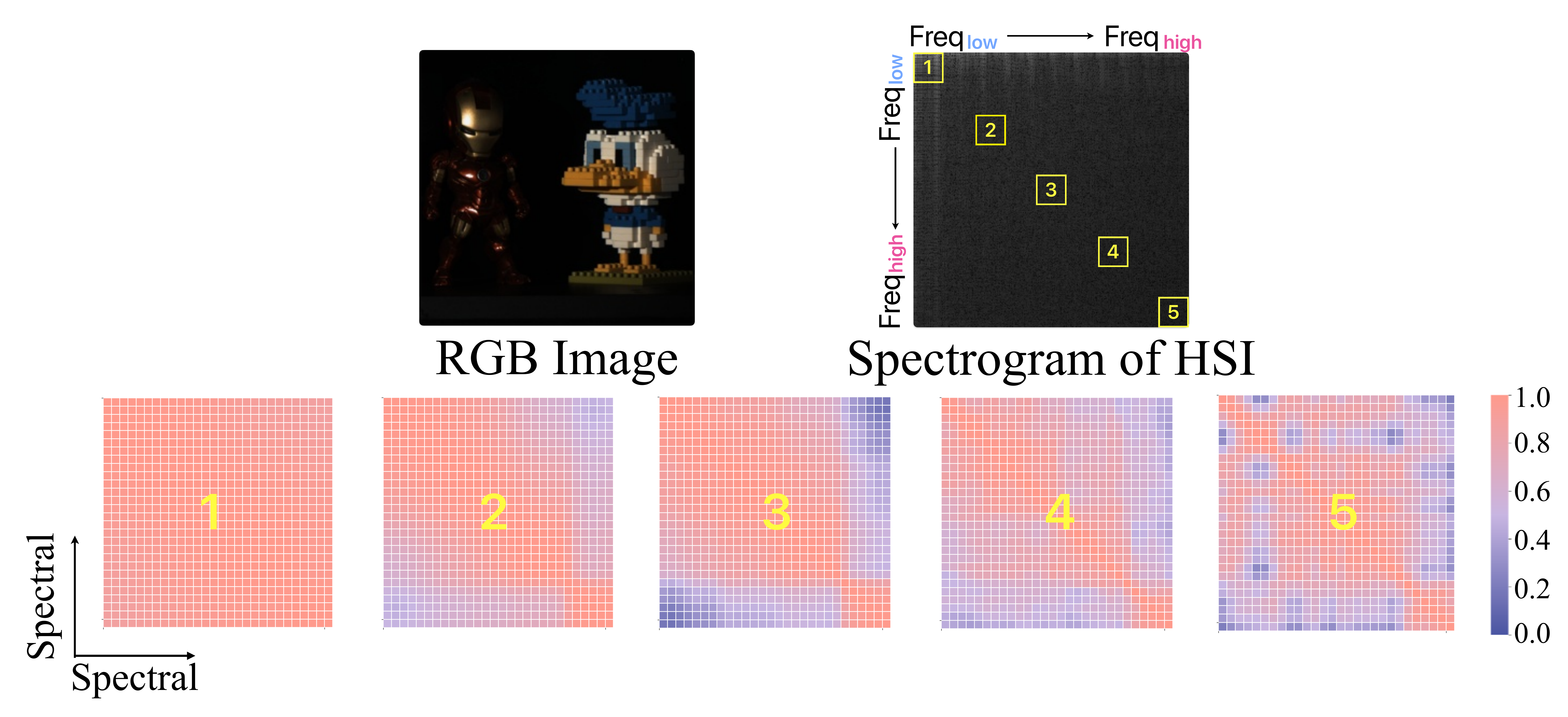}
        \caption{\small HFC in various patches}
        \label{fig:subfig2}
    \end{subfigure}
   \end{center}
   \vspace{-9mm}
   \caption{\small Visualization of the spectral-spatial correlation heatmap from frequency token-1 to token-5 in spectrogram of HSI.}
   \vspace{-8mm}
   \label{fig:HFC}
\end{figure}

\noindent\textbf{Spectral-spatial Correlation between Frequency Tokens.} The spectral-spatial correlation between frequency tokens is fundamentally inspired by the evident visual similarities across diverse spectra in the spectrograms of HSIs. To verify the correctness of the specific correlation, we collect 24 HSI datasets~\cite{data1,data2,data3,data4,data5,data6,data7,data9,data11,data12,data14,data15,data16,data17,data18,data20,data21,data22,data23}, which contains 1029 HSIs with divergent spatial size and spectral bands to conduct statistical analyses on the correlations between spectral-spatial token pairs.  We choose the Pearson-product-momentum correlation coefficient~\cite{pearson} as the metric, thereby sidestepping the influence of the numerical data scale.

\textbf{Firstly}, as shown in Fig.~\ref{fig:subfig1}, by computing the correlation for the HSI data in two domains, we derive two correlation maps of size $C\times C$, where $C$ is the number of spectral bands of HSI. Each element within the map represents the correlation degree of two spatial-vectorized spectral pairs with size $1\times HW$, where $H,W$ are the height and width of HSI. We further calculate the average value of all $C^2$ elements and obtain the average correlation.
\textbf{Subsequently}, we present comprehensive statistics of spectral correlation in two domains of 1029 HSIs including histograms and probability distribution. It is observed that the frequency domain spectral correlation is concentrated around 0.94, whereas the space domain counterpart is more dispersed and of a significant probability around 0.4, spotlighting the high stability and concentrated probability distribution of spectral-spatial correlation in the frequency domain.
\textbf{Additionally}, to unearth spectral correlation within specific frequency districts, we partition the HSI spectrogram into fixed-size spectral-spatial tokens and calculate the frequency spectral correlation between tokens across spectral bands. It should be emphasized that each frequency token embodies similar frequencies. Fig.~\ref{fig:subfig2} shows the descending trend of correlation from low to high-frequency tokens. 

Distilling our analyses, the spectral-spatial correlation between frequency tokens can be concisely summarized as:

\textbf{(i)} The frequency spectral-spatial token pairs unveil a high spectral correlation marked by significant stability, which remains relatively immune to distant spectra when juxtaposed with the space counterpart.

\textbf{(ii)} As the frequency ascends, there is a descending trend of the correlation between frequency tokens.

\begin{figure*}[t]
	\begin{center}
		\begin{tabular}[t]{c} \hspace{-2.4mm}
        \includegraphics[width=\textwidth]{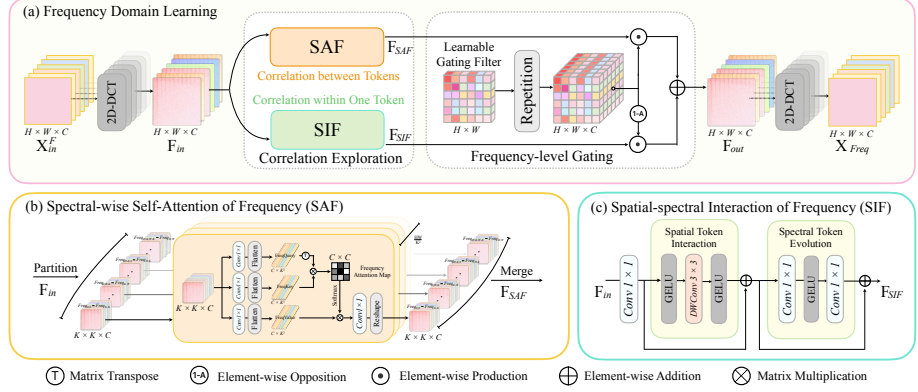}
		\end{tabular}
	\end{center}
	\vspace*{-7mm}
	\caption{\small The pipeline of the frequency domain learning. (a) The structure of the spectral-wise self-attention of frequency. (b) The structure of the spectral-spatial interaction of frequency.}
	\label{fig:FDLB}
	\vspace{-5mm}
\end{figure*}

The above two-fold correlations constitute the HFC prior. In the next section, we will leverage the HFC prior to establish the efficient frequency domain learning for comprehensive frequency exploitation.



\vspace{-1em}
\subsection{Frequency Domain Learning}
\vspace{-0.5em}

Inspired by the HFC prior, we formulate a novel frequency domain learning to exploit the image frequency information, which is composed of a correlation exploration block and a frequency-level gating block. The former consists of a Spectral-wise self-Attention in Frequency domain (SAF) and a Spectral-spatial Interaction in Frequency domain (SIF) to achieve thorough frequency exploitation. The latter provides dynamic gating capability to achieve the optimal balance between SAF and SIF.

\noindent\textbf{Spectral-wise self-Attention of Frequency.} The input of frequency domain learning ${\mathbf{X}_{in}^F}\in \mathbb{R}^{{H}\times{W}\times{C}}$ is initially transformed into frequency cube ${\mathbf{F}_{in}} \in \mathbb{R}^{H\times W\times C}$ via $C$ 2D-DCTs. For the low-frequency tokens, given the pronounced spectral-spatial correlation between frequency tokens, we creatively incorporate HFC with self-attention mechanism, sculpting the SAF to model the correlation between distant spectral-spatial frequency token pairs. As shown in Fig.~\ref{fig:FDLB}, in SAF, the input spectrogram ${\mathbf{F}_{in}}$ is firstly split into non-overlapping cubic patches with a size of $K\times K\times C$, denoted as [$f_1,...,f_n$], where $n=\frac{HW}{K^2}$. For each cube $f_i$, we map $f_i$ into FreqQuery ${\mathbf{Q}_i^F=\mathbf{W}_q^F\times f_i}$, FreqKey ${\mathbf{K}_i^F=\mathbf{W}_k^F\times f_i}$, FreqValue ${\mathbf{V}_i^F=\mathbf{W}_v^F\times f_i} \in \mathbb{R}^{K^2\times C}$ via three learnable parameters $\mathbf{W}_q^F, \mathbf{W}_k^F, \mathbf{W}_v^F \in \mathbb{R}^{C\times C}$.
Then the output of SAF dubbed $\mathbf{F}_{SAF} \in \mathbb{R}^{H\times W\times C}$ is calculated as:
\vspace{-0.6em}
\begin{equation}
    {\mathbf{F}_{i}}=\mathbf{V}_i^F {Softmax} (\frac{{\mathbf{Q}_i^F}^T \mathbf{K}_i^F}{\sqrt{C}}+\mathbf{P}_i^F),
\label{eq:pfcsa_attn1}
\end{equation}
\vspace{-1.5em}
\begin{equation}
    {\mathbf{F}_{SAF}}=\mathop{{Concat}}\limits_{i=1}^{n}(Reshape(Conv(\mathbf{F}_{i}))),
\label{eq:pfcsa_attn2}
\end{equation}
where $\mathbf{P}_i^F \in \mathbb{R}^{C\times C}$ are learnable parameters embedding the frequency position information. $\mathbf{F}_i \in \mathbb{R}^{K^2\times C}$ is output cubic of SAF. $Reshape(\cdot)$ transforms matrix $\mathbf{F}_i \in \mathbb{R}^{K^2\times C}$ into tensor $\mathbf{F}_i \in \mathbb{R}^{K\times K\times C}$.
In the implementation, the multi-head strategy is adopted to map features and exploit exhaustive information.

\begin{figure*}[t]
	\begin{center}
		\begin{tabular}[t]{c} \hspace{-2.4mm}
			\includegraphics[width=\textwidth]{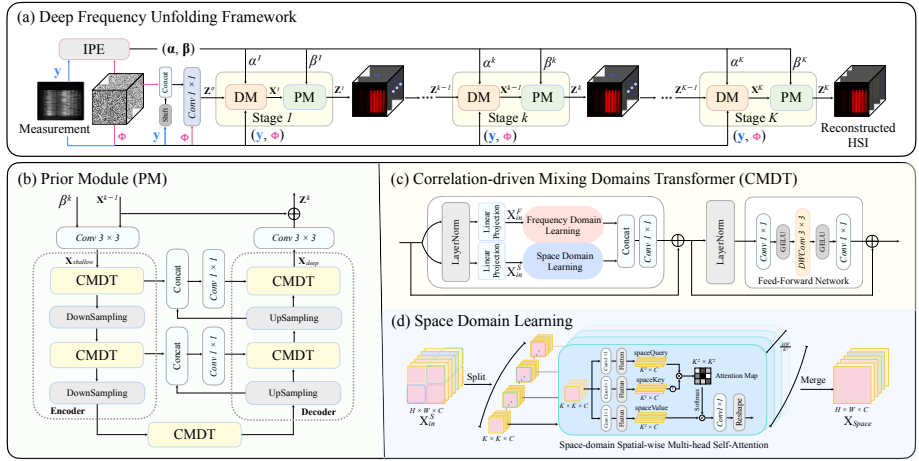}
		\end{tabular}
	\end{center}
	\vspace*{-7mm}
	\caption{\small Diagram of the overall framework. (a) Correlation-driven Mixing Domains Transformer-based Unfolding Framework. (b) The pipeline of the U-shaped prior module. (c) The correlation-driven mixing domains transformer. (d) The structure of space domain learning.}
	\label{fig:DST-DUF}
	\vspace{-3mm}
\end{figure*}

\noindent\textbf{Spectral-spatial Interaction of Frequency.} For the high-frequency tokens, considering the relation between correlation and self-attention, the learning of less related high frequencies is inefficient. Therefore, spectral-spatial interaction of frequency (SIF) is designed to accurately model high frequencies with spectral/spatial correlation and frequency inductive biases. More concretely, the inductive bias of the spatial token interaction is the same phase distance between neighboured frequencies. The inductive bias of the spectral token evolution is the same structural representation of one frequency across different spectral bands. As shown in Fig.~\ref{fig:FDLB}, The former employs the depth-wise $conv3\times3$ layer ${DW(\cdot)}$ with a number the same as the input channels, thus building a bridge for frequency tokens in the spatial neighborhood to interact with each other. The latter leverages two $conv1\times1$ layers ${Conv(\cdot)}$, thus facilitating a single frequency to evolve across the spectral groups. The outputs of two blocks are $\mathbf{F}_{spat}$ and $\mathbf{F}_{spec} \in \mathbb{R}^{H\times {W}\times C}$, respectively, thus effectively representing the frequency-level information in HSIs across both spatial and spectral dimensions. ${\sigma}$ represents the GELU function. The complete SIF process is represented as follows:
\begin{equation}
    {\mathbf{F}_{spat}}=\sigma \cdot DW (\sigma \cdot Conv(\mathbf{F}_{in}^i)),
\label{eq:sffm1}
\end{equation}
\vspace{-1.5em}
\begin{equation}
    {\mathbf{F}_{spec}}=Conv(\sigma \cdot Conv(\mathbf{F}_{spat}+\mathbf{F}_{in})),
\label{eq:sffm2}
\end{equation}
\vspace{-1.2em}
\begin{equation}
    {\mathbf{F}_{SIF}}=\mathbf{F}_{spec}+\mathbf{F}_{spat}+\mathbf{F}_{in}.
\label{eq:sffm3}
\end{equation}

\noindent\textbf{Learnable Gating Filter.} Considering the dominant correlation varying from low-frequency to high-frequency tokens, a learnable gating filter $\mathbf{LGF}$ with size $H\times {W}$ is introduced to adaptively determining the proportion of SAF and SIF, thereby ensuring the network to emphasize all frequencies. The output ${\mathbf{F}_{out}} \in \mathbb{R}^{H\times {W}\times C}$ is obtained as:
\begin{equation}
    \mathbf{F}_{out}={REP}(\mathbf{LGF})\odot \mathbf{F}_{SAF}+{REP}(1-\mathbf{LGF})\odot \mathbf{F}_{SIF}
\label{eq:fdlb}
\end{equation}
where ${REP(\cdot)}$ is the repetition operation along the channel dimension. Subsequently, we transform the ${\mathbf{F}_{out}}$ into space image ${\mathbf{X}_{Freq}} \in \mathbb{R}^{H\times {W}\times C}$ through C inverse 2D-DCTs, serving as the output of the frequency domain learning. 

\vspace{-1em}
\subsection{Correlation-driven Mixing Domains Transformer}
\vspace{-0.5em}
To exploit the pixel-level information as complementary to frequency domain learning, the space domain learning consisting of space-domain spatial-wise multi-head self-attention is introduced to capture image local correlation. As depicted in Fig.~\ref{fig:DST-DUF}, the input image ${\mathbf{X}_{in}^S}\in \mathbb{R}^{{H}\times{W}\times{C}}$ is also split into non-overlapping spatial-spectral tokens with size of $K\times K\times C$, denoted as [${x_1,...,x_n}$], where $n=\frac{HW}{K^2}$. The output of space domain learning dubbed $\mathbf{X}_{Space} \in \mathbb{R}^{H\times W\times C}$ is obtained after the operation.
Mixing the proposed frequency domain learning and the space domain learning, the CMDT based on traditional transformers is architected as shown in Fig.~\ref{fig:DST-DUF}. Concretely, the CMDT comprises layer normalization, linear projection, mixing domains learning comprising the frequency domain learning and the space domain learning, and feed-forward network, thus ensuring the comprehensive exploitation of HSI correlation in dual domains.



\subsection{Deep Frequency Unfolding Framework}
For convenience, given the sensing matrix $\mathbf{\Phi}$, the input HSI $\mathbf{x}$, and the imaging noise $\mathbf{n}$, the measurement $\mathbf{y}$ of CASSI imaging model can be vectorized as:
\begin{equation}
	\mathbf{y} = \mathbf{\Phi} \mathbf{x} + \mathbf{n}.
\label{eq:cassi}
\end{equation}

As~\cite{unfolding, yuanUnfolding} illustrates, the problem of reconstructing $\mathbf{x}$ from $\mathbf{y}$ can be split into two iterative process as:
\begin{equation}
	\mathbf{x}^{k+1} = \mathbf{z}^{k} + \mathbf{\Phi^T} [(\mathbf{y} - \mathbf{\Phi} \mathbf{z}^{k})./({\alpha + \mathbf{\Phi}\mathbf{\Phi^T}})],
\label{eq:dm}
\end{equation}
\vspace{-1em}
\begin{equation}
	\mathbf{z}^{k+1} = \mathbf{P}(\mathbf{x}^{k+1}, {\beta}),
\label{eq:pm}
\end{equation}
where variable $\mathbf{z}$ serves as an auxiliary variable, ${\mathbf{P}(\cdot)}$ refers to the deep prior network, $\alpha$ and $\beta$ are the iterative parameters. Detailed illustration of CASSI imaging model and the derivation of the unfolding framework can be found in ~\cite{dauhst,wangprior,roth2005fields}. Please note that the middle and final outputs of our method are in formulation of 3D cubes,~\emph{i.e.}, $\mathbf{X}_k$ and $\mathbf{Z}_k\in \mathbb{R}^{H\times W\times C}$, which can be obtained from the vectorized $\mathbf{x}^{k}$ and $\mathbf{z}^{k}$ by reshape operation.

By combining our CMDT with unfolding strategies, we architect the deep frequency unfolding framework with $K$ stages. As Fig.~\ref{fig:DST-DUF} shows, each stage contains a Data Module (DM) and a Prior Module (PM), corresponding to the calculation of the closed-form solution (Eq.~\ref{eq:dm}) and the deep prior exploitation (Eq.~\ref{eq:pm}). An Iteration Parameter Estimator (IPE)~\cite{dauhst} is employed to explore the degradation patterns and ill-posedness degree caused by the mask-modulation and dispersion-integration, thus providing iteration parameters $\alpha$ and $\beta$.

\noindent\textbf{Loss Function.} Given the ground-truth HSI $\bf{X}_{GT}$ and the predicted HSI $\bf{Z}_{K}\in \mathbb{R}^{H\times W\times C}$, the formula of the loss function is formulated as follows:
\begin{equation}
	Loss({\bf{X}}_{GT}, {\bf{Z}}_{K}) = ||{\bf{X}}_{GT}-{\bf{Z}}_{K}||_2,
\end{equation}
please note that the ${\bf{X}}_{GT}$ and ${\bf{Z}}_{K} \in \mathbb{R}^{H\times W\times C}$ are both space-domain HSI, that is to say, the loss function is calculated only in the space domain because the input and output of the overall pipeline are space-domain HSI.

\section{Experiments}
\label{sec:experiments}



\begin{table*}[t]
        \caption{The PSNR (top), SSIM (middle), and FDG (bottom) results of methods on $10$ scenes in KAIST.}
        \vspace{-2mm}
	\renewcommand{\arraystretch}{1.0}
	\newcommand{\tabincell}[2]{\begin{tabular}{@{}#1@{}}#2\end{tabular}}
	\centering
	\resizebox{0.98\textwidth}{!}
	{
		\centering
		\begin{tabular}{cccccccccccccc}
			\toprule[0.2em]
                \rowcolor{lightgray}
			~~~~Algorithms~~~~
                & ~Params$(M)$~
			& ~FLOPs$(G)$~
			& ~~~~S1~~~~
			& ~~~~S2~~~~
			& ~~~~S3~~~~
			& ~~~~S4~~~~
			& ~~~~S5~~~~
			& ~~~~S6~~~~
			& ~~~~S7~~~~
			& ~~~~S8~~~~
			& ~~~~S9~~~~
			& ~~~~S10~~~~
			& ~~~~Avg~~~~
			\\
			\midrule
			TwIST~\cite{twist}
                &\tabincell{c}{-}
			&\tabincell{c}{-}
			&\tabincell{c}{25.16\\0.700\\20.38}
			&\tabincell{c}{23.02\\0.604\\26.25}
			&\tabincell{c}{21.40\\0.711\\32.83}
			&\tabincell{c}{30.19\\0.851\\12.42}
			&\tabincell{c}{21.41\\0.635\\29.93}
			&\tabincell{c}{20.95\\0.644\\31.72}
			&\tabincell{c}{22.20\\0.643\\28.65}
			&\tabincell{c}{21.82\\0.650\\30.69}
			&\tabincell{c}{22.42\\0.690\\34.26}
			&\tabincell{c}{22.67\\0.569\\35.26}
			&\tabincell{c}{23.12\\0.669\\28.24}
			\\
			\midrule
			GAP-TV~\cite{gaptv}
                &\tabincell{c}{-}
			&\tabincell{c}{-}
			&\tabincell{c}{26.82\\0.754\\17.60}
			&\tabincell{c}{22.89\\0.610\\19.48}
			&\tabincell{c}{26.31\\0.802\\16.01}
			&\tabincell{c}{30.65\\0.852\\5.14}
			&\tabincell{c}{23.64\\0.703\\21.58}
			&\tabincell{c}{21.85\\0.663\\25.07}
			&\tabincell{c}{23.76\\0.688\\23.71}
			&\tabincell{c}{21.98\\0.655\\28.94}
			&\tabincell{c}{22.63\\0.682\\23.66}
			&\tabincell{c}{23.10\\0.584\\26.19}
			&\tabincell{c}{24.36\\0.669\\20.74}
			\\
                \midrule
			GAP-Net ~\cite{gapnet}
                &\tabincell{c}{4.27}
			&\tabincell{c}{78.58}
			&\tabincell{c}{33.74\\0.911\\11.82}
			&\tabincell{c}{33.26\\0.900\\11.11}
			&\tabincell{c}{34.28\\0.929\\11.70}
			&\tabincell{c}{41.03\\0.967\\6.24}
			&\tabincell{c}{31.44\\0.919\\14.03}
			&\tabincell{c}{32.40\\0.925\\11.66}
			&\tabincell{c}{32.27\\0.902\\14.27}
			&\tabincell{c}{30.46\\0.905\\14.81}
			&\tabincell{c}{33.51\\0.915\\13.05}
			&\tabincell{c}{30.24\\0.895\\16.07}
			&\tabincell{c}{33.26\\0.917\\12.47}
			\\
                \midrule
			DGSMP ~\cite{dgsmp}
                &\tabincell{c}{3.76}
			&\tabincell{c}{646.65}
			&\tabincell{c}{33.26\\0.915\\12.64}
			&\tabincell{c}{32.09\\0.898\\13.67}
			&\tabincell{c}{33.06\\0.925\\16.81}
			&\tabincell{c}{40.54\\0.964\\8.49}
			&\tabincell{c}{28.86\\0.882\\19.07}
			&\tabincell{c}{33.08\\0.937\\11.97}
			&\tabincell{c}{30.74\\0.886\\16.60}
			&\tabincell{c}{31.55\\0.923\\13.94}
			&\tabincell{c}{31.66\\0.911\\17.62}
			&\tabincell{c}{31.44\\0.925\\14.37}
			&\tabincell{c}{32.63\\0.917\\14.52}
			\\
                \midrule
                $\lambda$-Net ~\cite{lambdanet}
                &\tabincell{c}{62.64}
			&\tabincell{c}{117.98}
			&\tabincell{c}{31.48\\0.858\\16.67}
			&\tabincell{c}{31.09\\0.842\\20.22}
			&\tabincell{c}{28.96\\0.823\\18.31}
			&\tabincell{c}{34.56\\0.902\\9.15}
			&\tabincell{c}{28.53\\0.808\\23.56}
			&\tabincell{c}{30.83\\0.877\\20.88}
			&\tabincell{c}{28.71\\0.824\\24.10}
			&\tabincell{c}{30.09\\0.881\\24.12}
			&\tabincell{c}{30.43\\0.868\\22.29}
			&\tabincell{c}{28.78\\0.842\\24.72}
			&\tabincell{c}{30.35\\0.852\\20.40}
			\\
                \midrule
			HDNet ~\cite{hdnet}
                &\tabincell{c}{2.37}
			&\tabincell{c}{154.76}
			&\tabincell{c}{35.14\\0.935\\10.08}
			&\tabincell{c}{35.67\\0.940\\8.67}
			&\tabincell{c}{36.03\\0.943\\9.88}
			&\tabincell{c}{42.30\\0.969\\4.68}
			&\tabincell{c}{32.69\\0.946\\12.70}
			&\tabincell{c}{34.46\\0.952\\9.36}
			&\tabincell{c}{33.67\\0.926\\12.17}
			&\tabincell{c}{32.48\\0.941\\12.04}
			&\tabincell{c}{34.89\\0.942\\12.02}
			&\tabincell{c}{32.38\\0.937\\13.42}
			&\tabincell{c}{34.97\\0.943\\10.50}
			\\
			\midrule
			MST-L ~\cite{mst}
                &\tabincell{c}{2.03}
			&\tabincell{c}{28.15}
			&\tabincell{c}{35.40\\0.941\\9.53}
			&\tabincell{c}{35.87\\0.944\\8.08}
			&\tabincell{c}{36.51\\0.953\\10.23}
			&\tabincell{c}{42.27\\0.973\\6.23}
			&\tabincell{c}{32.77\\0.947\\12.34}
			&\tabincell{c}{34.80\\0.955\\9.18}
			&\tabincell{c}{33.66\\0.925\\11.39}
			&\tabincell{c}{32.67\\0.948\\11.61}
			&\tabincell{c}{35.39\\0.949\\12.27}
			&\tabincell{c}{32.50\\0.941\\12.61}
			&\tabincell{c}{35.18\\0.948\\10.35}
			\\
			\midrule
			CST-L ~\cite{cst}
                &\tabincell{c}{3.00}
			&\tabincell{c}{40.01}
			&\tabincell{c}{35.96\\0.949\\9.60}
			&\tabincell{c}{36.84\\0.955\\7.81}
			&\tabincell{c}{38.16\\0.962\\9.00}
			&\tabincell{c}{42.44\\0.975\\5.64}
			&\tabincell{c}{33.25\\0.955\\11.86}
			&\tabincell{c}{35.72\\0.963\\8.28}
			&\tabincell{c}{34.86\\0.944\\11.16}
			&\tabincell{c}{34.34\\0.961\\10.54}
			&\tabincell{c}{36.51\\0.957\\11.10}
			&\tabincell{c}{33.09\\0.945\\11.97}
			&\tabincell{c}{36.12\\0.957\\9.70}
			\\
			\midrule
			DAUHST-9stg ~\cite{dauhst}
                &\tabincell{c}{6.15}
			&\tabincell{c}{79.50}
			&\tabincell{c}{37.25\\0.958\\7.37}
			&\tabincell{c}{39.02\\0.967\\5.64}
			&\tabincell{c}{41.05\\0.971\\4.84}
			&\tabincell{c}{46.15\\0.983\\2.69}
			&\tabincell{c}{35.80\\0.969\\8.48}
			&\tabincell{c}{37.08\\0.970\\6.91}
			&\tabincell{c}{37.57\\0.963\\7.04}
			&\tabincell{c}{35.10\\0.966\\8.81}
			&\tabincell{c}{40.02\\0.970\\5.90}
			&\tabincell{c}{34.59\\0.956\\9.86}
			&\tabincell{c}{38.36\\0.967\\6.75}
			\\
                \midrule
                \rowcolor{rouse}
			\bf{Ours-2stg}
                &\tabincell{c}{\bf0.90}
			&\tabincell{c}{\bf21.38}
                &\tabincell{c}{36.49\\0.954\\7.94}
			&\tabincell{c}{38.09\\0.960\\6.18}
			&\tabincell{c}{39.87\\0.968\\5.46}
			&\tabincell{c}{45.29\\0.984\\3.06}
			&\tabincell{c}{35.26\\0.965\\8.94}
			&\tabincell{c}{36.16\\0.966\\7.70}
			&\tabincell{c}{36.20\\0.955\\8.42}
			&\tabincell{c}{34.36\\0.959\\9.56}
			&\tabincell{c}{38.90\\0.967\\6.42}
			&\tabincell{c}{33.74\\0.951\\10.54}
			&\tabincell{c}{37.44\\0.963\\7.42}
                \\
                \midrule
                \rowcolor{rouse}
                \bf{Ours-3stg}
                &\tabincell{c}{\bf0.90}
			&\tabincell{c}{31.56}
                &\tabincell{c}{37.16\\0.958\\7.41}
			&\tabincell{c}{38.49\\0.962\\5.99}
			&\tabincell{c}{41.03\\0.973\\4.64}
			&\tabincell{c}{46.01\\0.985\\2.71}
			&\tabincell{c}{36.01\\0.970\\8.33}
			&\tabincell{c}{36.73\\0.968\\7.18}
			&\tabincell{c}{37.00\\0.961\\7.65}
			&\tabincell{c}{34.85\\0.960\\8.98}
			&\tabincell{c}{39.71\\0.971\\5.74}
			&\tabincell{c}{34.59\\0.957\\9.30}
			&\tabincell{c}{38.16\\0.967\\6.80}
                \\
                \midrule
                \rowcolor{rouse}
                \bf{Ours-5stg}
                &\tabincell{c}{\bf0.90}
			&\tabincell{c}{51.90}
                &\tabincell{c}{37.57\\0.964\\6.78}
			&\tabincell{c}{39.89\\0.975\\5.04}
			&\tabincell{c}{42.10\\0.978\\4.02}
			&\tabincell{c}{46.98\\0.990\\2.35}
			&\tabincell{c}{36.90\\0.975\\7.24}
			&\tabincell{c}{{37.10}\\0.975\\6.90}
			&\tabincell{c}{37.99\\0.968\\6.64}
			&\tabincell{c}{{35.11}\\{0.969}\\8.64}
			&\tabincell{c}{40.91\\0.978\\5.01}
			&\tabincell{c}{34.95\\0.964\\8.85}
			&\tabincell{c}{38.96\\0.974\\6.15}
                \\
                \midrule
                \rowcolor{rouse}
                \bf{Ours-9stg}
                &\tabincell{c}{\bf0.90}
			&\tabincell{c}{92.59}
                &\tabincell{c}{\bf{37.88}\\\bf{0.968}\\\bf6.46}
			&\tabincell{c}{\bf{40.50}\\\bf{0.979}\\\bf4.69}
			&\tabincell{c}{\bf{42.34}\\\bf{0.979}\\\bf3.86}
			&\tabincell{c}{\bf{47.76}\\\bf{0.992}\\\bf2.16}
			&\tabincell{c}{\bf{37.70}\\\bf{0.979}\\\bf6.56}
			&\tabincell{c}{\bf{37.49}\\\bf{0.977}\\\bf6.58}
			&\tabincell{c}{\bf{38.19}\\\bf{0.970}\\\bf6.42}
			&\tabincell{c}{\bf{35.37}\\\bf{0.971}\\\bf8.38}
			&\tabincell{c}{\bf{42.37}\\\bf{0.982}\\\bf4.01}
			&\tabincell{c}{\bf{35.05}\\\bf{0.966}\\\bf8.75}
			&\tabincell{c}{\bf{39.47}\\\bf{0.976}\\\bf5.79}
                \\
			\bottomrule[0.2em]
		\end{tabular}
    }
	\vspace{-5mm}
	\label{tab:performance}
\end{table*}

\subsection{Experimental Setup}
\noindent\textbf{Datasets.} We adopt CAVE~\cite{data3} and KAIST~\cite{data5} with 28 wavelengths from 450nm to 650nm for experiments. CAVE consists of 32 HSIs with spatial size $512\times512$. KAIST contains 30 HSIs of spatial size $2704\times3376$. We use the augmented CAVE for training, 10 scenes from KAIST for simulation testing, and 5 scenes from ~\cite{tsanet} for real testing.

\begin{figure*}[t]
	\begin{center}
		\begin{tabular}[t]{c} \hspace{-2.4mm}
			\includegraphics[width=0.98\textwidth]{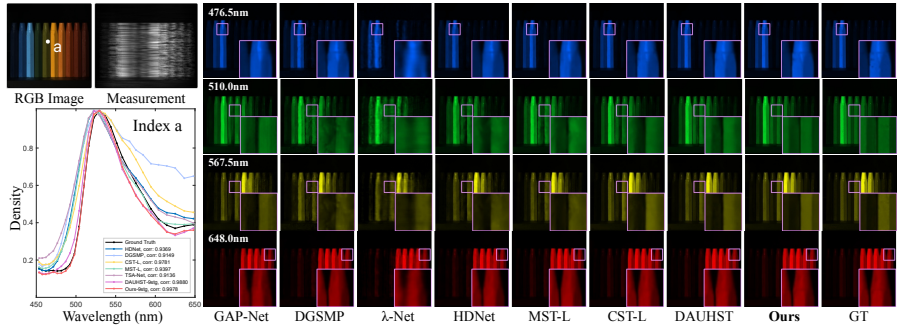}
		\end{tabular}
	\end{center}
	\vspace*{-7mm}
	\caption{\small Visual quality comparisons of simulation experiment. Left: the spectral density curves. Right: perceptual quality with 4 out of 28 spectral bands.}
	\label{fig:simu}
	\vspace{-2mm}
\end{figure*}

\begin{figure*}[t]
	\begin{center}
		\begin{tabular}[t]{c} \hspace{-2.4mm}
			\includegraphics[width=0.98\textwidth]{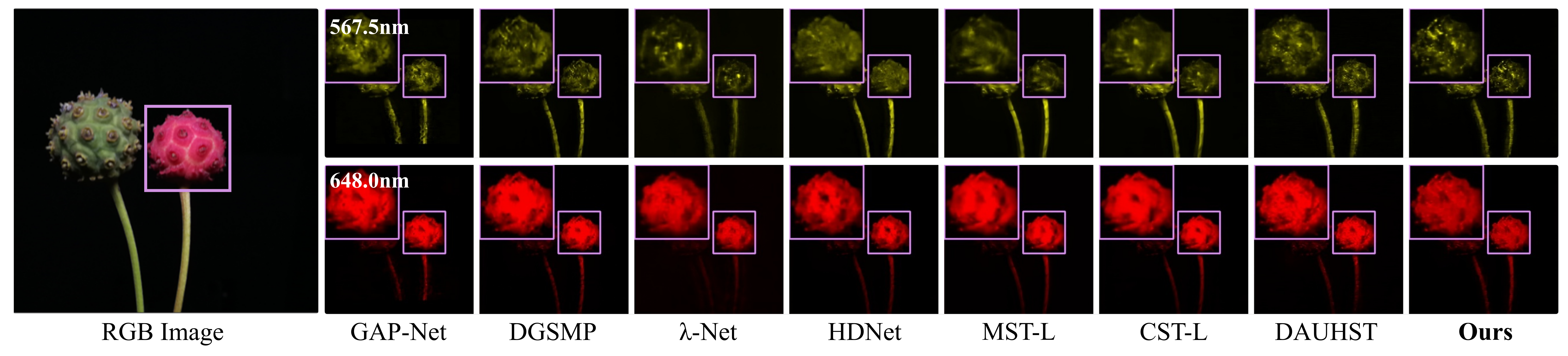}
		\end{tabular}
	\end{center}
	\vspace*{-7mm}
	\caption{\small Visual quality comparisons of real experiment.}
	\label{fig:real}
	\vspace{-5mm}
\end{figure*}

\noindent\textbf{Implementation Details.} We implement our method by PyTorch. All methods are trained with Adam optimizer ($\beta_1=0.9, \beta_2=0.999$) using the Cosine Annealing scheme for 300 epochs on an RTX 3090 GPU. The initial learning rate is $4\times 10^{-4}$, batch size is 5. Some data augmentations including random rotation and flipping are used to enrich the diversity of training data. The kernel size of SAF is set to 8. The parameters of PM are shared between different stages. The training objective is to minimize the Root Mean Square Error (RMSE) between reconstructed and ground-truth (GT) HSIs. The source codes and pre-trained models will be released to be publicly available.

\noindent\textbf{Compared Methods.} We compare our method with two traditional methods (TwIST~\cite{twist} and GAP-TV), five end-to-end methods ({$\lambda$-Net}, GAP-Net~\cite{gaptv}, HDNet~\cite{hdnet}, MST~\cite{mst}, and CST~\cite{cst}) and four deep unfolding methods (DGSMP~\cite{dgsmp}, GAP-Net~\cite{gapnet}, DAUHST~\cite{dauhst}, and RDLUF~\cite{rdluf}). Given that RDLUF changes the sensing matrix that should have been fixed, the direct comparison with RDLUF is unfair. Thus, we conduct a further combination study for a fair comparison with RDLUF in the learnable sensing matrix part.

\noindent\textbf{Evaluation Metrics.} The reconstruction quality is evaluated by perceptual quality and numerical performance including PSNR, SSIM, and Frequency Domain Gap (FDG). The best numerical value in each metric is in bold form.
The calculation of FDG is in the supplementary material.
\vspace{-1em}

\begin{wraptable}{r}{0.47\textwidth}
    \vspace{-12mm}
    \centering
    \caption{\small Noise injection experiment.}
    \vspace{2mm}
    \scalebox{0.70}{
        \begin{tabular}{c c c  c c c c}
        		\toprule
                    \rowcolor{lightgray}
        		Sigma & Metric & HDNet & DAUHST & Ours \\
        		\midrule
        		\multirow{3}{*}{$0.01$} & PSNR & 21.81 & 35.15 & \bf36.34 \\
                    \multirow{3}{*}{} & SSIM & 0.558 & 0.390 & \bf0.951 \\
                    \multirow{3}{*}{} & Preserve Degree & 63.51\% & 96.72\% & \bf97.42\% \\
                    \midrule
        		\multirow{3}{*}{$0.05$} & PSNR & 15.13 & 27.06 & \bf31.40 \\
                    \multirow{3}{*}{} & SSIM & 0.208 & 0.726 & \bf0.865 \\
                    \multirow{3}{*}{} & Preserve Degree & 44.05\% & 74.46\% & \bf84.18\% \\
                    \midrule
                    \multirow{3}{*}{$0.1$} & PSNR & 12.20 & 24.22 & \bf27.30 \\
                    \multirow{3}{*}{} & SSIM & 0.093 & 0.632 & \bf0.731 \\
                    \multirow{3}{*}{} & Preserve Degree & 35.52\% & 66.64\% & \bf73.19\% \\
        		\bottomrule
        \end{tabular}
    }
    \label{tab:table2}
    \vspace{-5mm}
\end{wraptable}
\subsection{Noise Injection Experiment}
\vspace{-0.2em}
To assess the noise robustness, we introduce a series of zero-mean Gaussian noise with a standard deviation ranging from 0.01 to 0.1 to measurements and conduct tests on pre-trained model. Considering the fairness of computational costs and characters of various models, DAUHST and HDNet are chosen for comparisons. Tab.~\ref{tab:table2} indicates the max preserving degree of our method, exhibiting the best robustness to noise.

\vspace{-1em}
\subsection{Simulation Results}
\noindent\textbf{Numerical Results.} The results from 10 simulated scenes are represented in Tab.~\ref{tab:performance}, we also provide the line graph in Fig.~\ref{fig:psnr-flops} for clear comparison, which shows that our model achieves the optimal balance between computational cost and performance. Compared to DAUHST, Ours-9stg achieves outstanding performance, $i.e.$, 39.47dB in PSNR, while Ours-5stg excels DAUHST-9stg with 0.6dB increase in PSNR and only requires 65$\%$ FLOPs and 15$\%$ Params, verifying the efficiency of our method.

\vspace{-2mm}
\hspace{-2em}
\begin{minipage}{\textwidth}
\vspace{-1mm}
 \begin{minipage}[t]{0.48\textwidth}
    \centering
     \makeatletter\def\@captype{table}\makeatother\caption{Break-down ablation.} 
     \vspace{-2mm}
     \label{tab:breakdown}
     \scalebox{0.6}{
			\begin{tabular}{c c c  c c c c}
				\toprule
				\rowcolor{color3} Baseline-1 &SAF &SIF   &PSNR &SSIM &Params (M) &FLOPS (G) \\
				\midrule
				\checkmark & & &36.02 &0.952 &0.55 &16.30 \\
				\checkmark  &\checkmark & &37.10 &0.960 &0.64 &20.06 \\
				\checkmark  &\checkmark &\checkmark &\bf 37.44 &\bf 0.963 &0.90 &21.39 \\
				\bottomrule
	\end{tabular}}
  \end{minipage}
  \hspace{1mm}
  \begin{minipage}[t]{0.48\textwidth}
   \centering
   \makeatletter\def\@captype{table}\makeatother
        \caption{Ablation of kernel size in SAF.}
        \vspace{-2mm}
        \label{tab:kernelsize}
        \scalebox{0.55}{
        \begin{tabular}{c c c c c c}
				\toprule
				\rowcolor{color3} Model &~Baseline-3~ &~kernel-2~ &~kernel-4~ &~kernel-8~ &~kernel-16~\\
				\midrule
				PSNR & 34.00 &34.25 &34.57 &\bf34.61 &34.28\\
				SSIM  &0.930 &0.934 &0.939 &\bf0.940 &0.936\\
				\bottomrule
	\end{tabular}}
   \end{minipage}
   \vspace{2mm}
\end{minipage}
\hspace{-1em}
\begin{minipage}{\textwidth}
 \begin{minipage}[t]{0.50\textwidth}
    \centering
    \makeatletter\def\@captype{table}\makeatother\caption{\small Ablation of various self-attention mechanisms.}
    \vspace{-2mm}
     \label{tab:selfattention}
     \scalebox{0.58}{
			\begin{tabular}{l c c c c c c}
				\toprule
				\rowcolor{color3} Method &Baseline-2 &G-MSA &SW-MSA &S-MSA &HS-MSA  &\bf SAF\\
				\midrule
				PSNR &32.79 &33.63 &33.75 &33.82 &34.05 &\bf 34.61 \\
				SSIM &0.904 &0.920 &0.924 &0.926 &0.930  &\bf 0.940\\
				Params (M)  &0.40 &0.48 &0.48 &0.48 &0.48  & 0.48\\
				FLOPS (G)  &6.85 &10.30 &9.41 &8.89     &9.72 & \bf7.87\\
                Complexity &- &$H^2W^2C$ &$2K^2HW$ &$HWC^2$ &$K^2HWC$ &\bf$HWC^2$\\
				\bottomrule
	\end{tabular}}
  \end{minipage}
  \begin{minipage}[t]{0.48\textwidth}
   \centering
   \makeatletter\def\@captype{table}\makeatother
        \caption{\small Ablation of parameter sharing.}
        \vspace{-2mm}
        \label{tab:sharing}
        \scalebox{0.62}{\begin{tabular}{c c c c c}
				\toprule
				\rowcolor{color3}stages &~Params (M)~ &~FLOPs (G)~ &~PSNR~ & ~SSIM~   \\
				\midrule
				2 &1.75   &21.39 &\bf37.64  &\bf 0.966  \\
				2-shared  &0.90    &21.39  &37.44 &0.963  \\
				3  &2.60    &31.56  &38.05  &0.967 \\
				3-shared &0.90    &31.56  &\bf38.16 &\bf0.968  \\
                5  &4.31    &51.90  &38.42  &0.971 \\
				5-shared  &0.90    &38.42  &\bf38.96 &\bf0.974  \\
				\bottomrule
	\end{tabular}}
   \end{minipage}
   \vspace{1mm}
\end{minipage}

\begin{wrapfigure}{t}{0.47\textwidth}
   \vspace{-14mm}
   \begin{center}
      \includegraphics[width=0.50\textwidth]{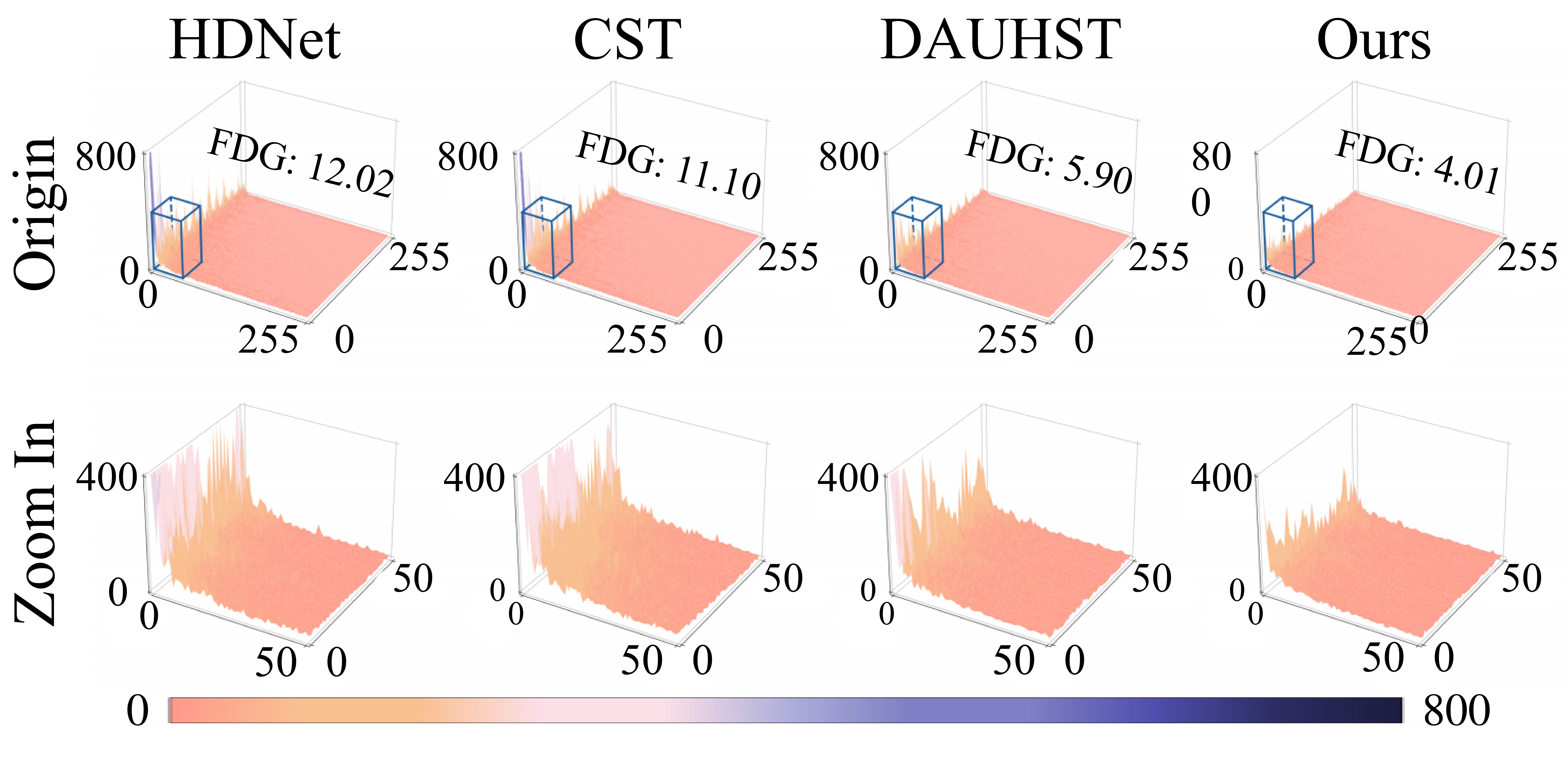}
   \end{center}
   \vspace{-14mm}
   \caption{\small Visualisation of HFC heatmaps.}
   \vspace{-9mm}
   \label{fig:FDLcomp}
\end{wrapfigure} 

\noindent\textbf{Frequency Domain Gap.} To show the efficacy of our method in learning image frequency information, we visualize the spectrogram residual maps of our method, HDNet, CST, and DAUHST to GT. As shown in Fig.~\ref{fig:FDLcomp}, our method exhibits the smoothest surface, especially in the zoomed area, and has the lowest FDG, thus demonstrating the superior recovery of the HSI frequencies.

\noindent\textbf{Perceptual Quality.} For better vision, we exhibit the visual results in RGB format with CIE color as the mapping function. The right part in Fig.~\ref{fig:simu} shows that our method excels in preserving clear edges, particularly in the zoomed area. The left part in Fig.~\ref{fig:simu} depicts the spectral curves at index $a$ in RGB image, the highest correlation and the closest curve to GT further emphasize the efficacy of our method.


\vspace{-1em}

\subsection{Real Experiment}
We further evaluate the effectiveness of our method on real HSI data.  Following the same settings as ~\cite{tsanet}, we retrain our method of 2 stages with the real mask on the CAVE and KAIST datasets. Fig.~\ref{fig:real} presents the visual comparison between our method and others. Compared to others, our method preserves more details with fewer artifacts, demonstrating the capability of our method to reconstruct the accurate details in HSIs. 

\vspace{-1em}
\subsection{Ablation Study}





\noindent\textbf{Break-down Ablation of Frequency Domain Learning.} To verify the effectiveness of each component of the frequency domain learning, we present the break-down ablation experiments on SAF and SIF. Baseline-1 is derived by removing SAF and SIF from our method on 2 stages. As Tab.~\ref{tab:breakdown} shows, the baseline-1 achieves 36.02 dB. When we apply SAF and SIM, the method achieves 1.08 dB, and 0.34 dB improvements, showing the effectiveness of each component.



\hspace{-1em}
\begin{minipage}{\textwidth}
 \begin{minipage}[t]{0.50\textwidth}
    \centering
    \makeatletter\def\@captype{table}\makeatother
    \caption{\small Fair comparison with RDLUF.}
    \label{tab:rdlufcomp}
    \vspace{-2mm}
    \scalebox{0.60}{
        \begin{tabular}{c c c  c c c c}
        		\toprule
                    \rowcolor{lightgray}
        		~~stage~~ & ~~Ours~~ & ~~RDLUF~~ & ~~Ours-plus~~\\
        		\midrule
                    PSNR & 38.16 & 37.56 & 38.20\\
                    SSIM & 0.967 & 0.963 & 0.972 \\
                    FLOPs (G) & 31.56 & 62.34 & 32.29 \\
                    Params (M) & 0.9 & 1.89 & 0.82 \\
        		\bottomrule
        \end{tabular}}
  \end{minipage}
  \hspace{-6mm}
  \begin{minipage}[t]{0.50\textwidth}
   \centering
   \makeatletter\def\@captype{table}\makeatother
        \caption{\small HFC metrics.}
        \label{tab:hfc1}
        \vspace{-2mm}
        \scalebox{0.7}{
        \begin{tabular}{c c c c}
        		\toprule
                    \rowcolor{lightgray}
        		metric &~~~~HFC~~~~ \\
        		\midrule
                Inference Time (s) &0.0592\\
                Computation Effort (GFLOPs) &0.498\\
                Memory Cost (M) &0.086\\
        		\bottomrule
        \end{tabular}}
   \end{minipage}
   \vspace{1mm}
\end{minipage}

\noindent\textbf{Ablation Study of Kernel Size in SAF.} To obtain the optimal kernel size for SAF, which balances the number of similar frequencies and the correlation between frequency tokens, we conduct the ablation experiments on SAF of various kernel sizes, \emph{e.g.}, 2, 4, 8, 16. Baseline-3 is adopted by removing SIF and space domain learning from CMDT with 1 stage. As Tab.~\ref{tab:kernelsize} illustrates, the kernel-8 model achieves the best performance of 34.61 dB in PSNR, thus adopted by us.

\noindent\textbf{Ablation Study of Self-attention Mechanisms.} 
To compare SAF with other self-attention mechanisms, we adopt G-MSA~\cite{gmsa}, Swin MSA (SW-MSA)~\cite{swmsa}, and HS-MSA~\cite{dauhst} as competitors. The baseline-2 is adopted by removing CMDT from our method of 1 stage. Tab.~\ref{tab:selfattention} shows that SAF yields the highest numerical performance with the linear complexity, demonstrating the prosperous performance of self-attention in the frequency domain.

\noindent\textbf{Ablation Study of Parameter Sharing.}
To validate the effects of parameter sharing, a series of experiments are conducted in various stages. As Tab.~\ref{tab:sharing} shows, in 2 stages, the method without parameter sharing yields higher performance than its counterpart. However, in long stages,~\emph{e.g.}, 3 stages and 5 stages, the method with parameter sharing excels its counterpart, which is primarily because methods without parameter sharing tend to experience multiplicative growth of parameters as the stage increases, leading to the overfitting of the neural network, especially in the situation where the data diversity is insufficient. Therefore, the method with parameter sharing exhibits promising reconstruction performance.

\noindent\textbf{Learnable sensing matrix.}
Since RDLUF changes the forward process with new learnable $\mathbf{\Phi}$ in each iteration, whose performance comparison with our method is fundamentally unfair. For a fair comparison, We further conduct the combination experiment to validate the superiority of our CMDT, $i.e.$, replace CMDT with Mix$S^2$ transformer in RDLUF and retrain the model on 3 stages. As shown in Tab.~\ref{tab:rdlufcomp} which shows that the proposed CMDT offers a 0.64 dB increase in PSNR and 0.09 increase in SSIM while requiring only 43.3$\%$ Params and 51.8$\%$ FLOPs compared to RDLUF.

\begin{wrapfigure}{t}{0.47\textwidth}
   \vspace{-19mm}
   \begin{center}
      \includegraphics[width=0.45\textwidth]{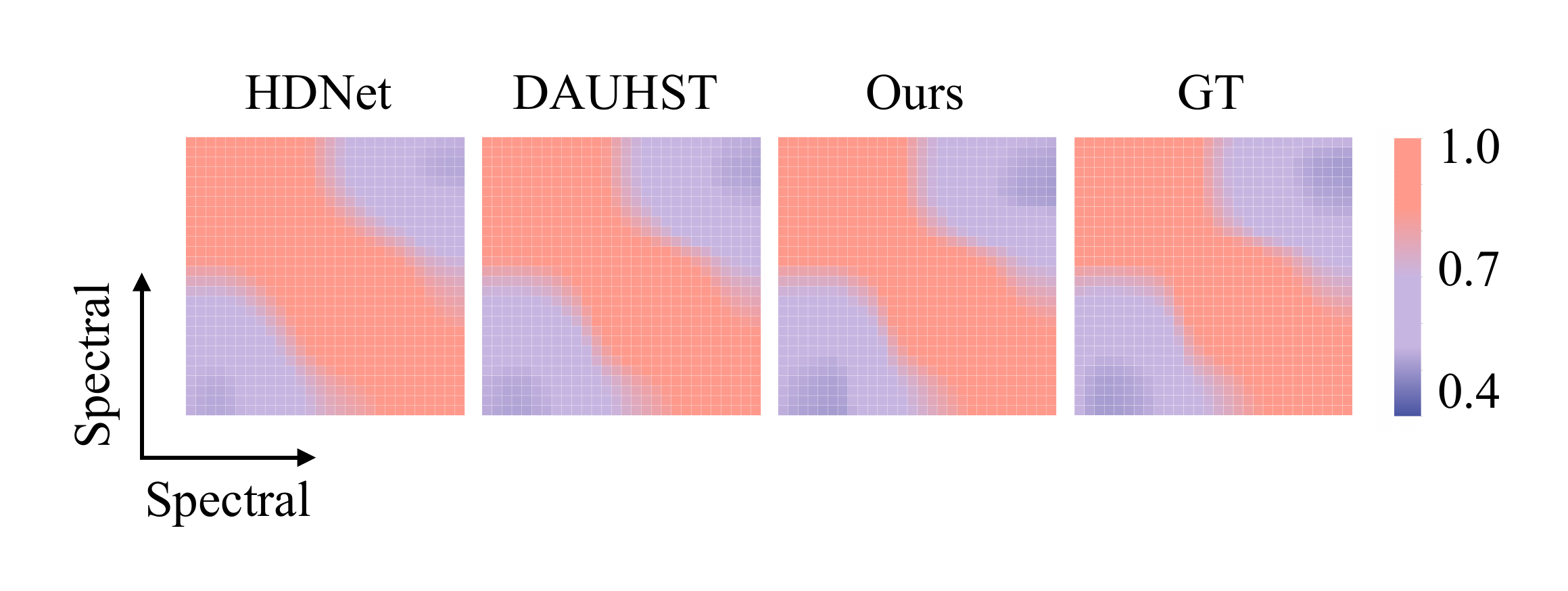}
   \end{center}
   \vspace{-16mm}
   \caption{\small Visualisation of HFC heatmaps.}
   \vspace{-10mm}
   \label{fig:subfig22}
\end{wrapfigure} 

\noindent\textbf{Revisiting HFC.} First, since our method is correlation-driven, we redraw the spectral correlation heatmap of reconstructed HSI in the frequency domain. HDNet and DAUHST are chosen for comparison. As Fig.~\ref{fig:subfig22} depicts, our method gives the closest spectral correlation reconstruction to GT, especially in the left-bottom district, validating the accurate correlation reconstruction of our method.
Second, to validate the running cost of our frequency domain learning network based on HFC, we compute inference time, computation effort, and memory cost in the network, showing the satisfactory performance in Tab.~\ref{tab:hfc1}.

\vspace{-1em}
\section{Discussion}
\label{discuss}
\begin{wrapfigure}{t}{0.45\textwidth}
   \vspace{-16mm}
   \begin{center}
      \includegraphics[width=0.45\textwidth]{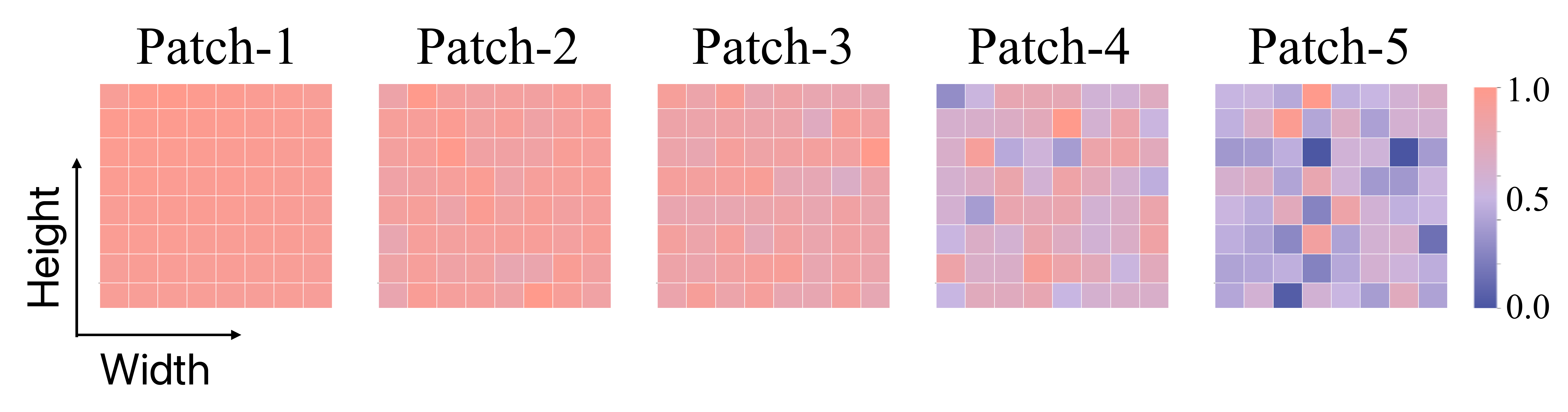}
   \end{center}
   \vspace{-14mm}
   \caption{\small Heatmaps of learnable frequency filter.}
   \vspace{-10mm}
   \label{fig:subfig11}
\end{wrapfigure} 

\vspace{-0.5em}
\subsection{Visualization of Learnable Gating Filter}
To prove the gating capacity of the learnable gating filter, we draw the heatmaps of 5 increasing-frequency tokens named patch-1 to patch-5. As Fig.~\ref{fig:subfig11} shows, the higher the frequency tokens are, the lower the overall values in the corresponding patch district of the learnable gating filter are, thus verifying that the learnable filter makes the low-frequency tokens with dominant spectral-spatial correlation mostly come from the SAF and high-frequency tokens with dominant spectral/spatial correlation mostly come from the SIF.

\vspace{-0.6em}
\subsection{Relation between attention map and correlation}
\begin{wrapfigure}{t}{0.50\textwidth}
   \vspace{-14mm}
   \begin{center}
      \includegraphics[width=0.50\textwidth]{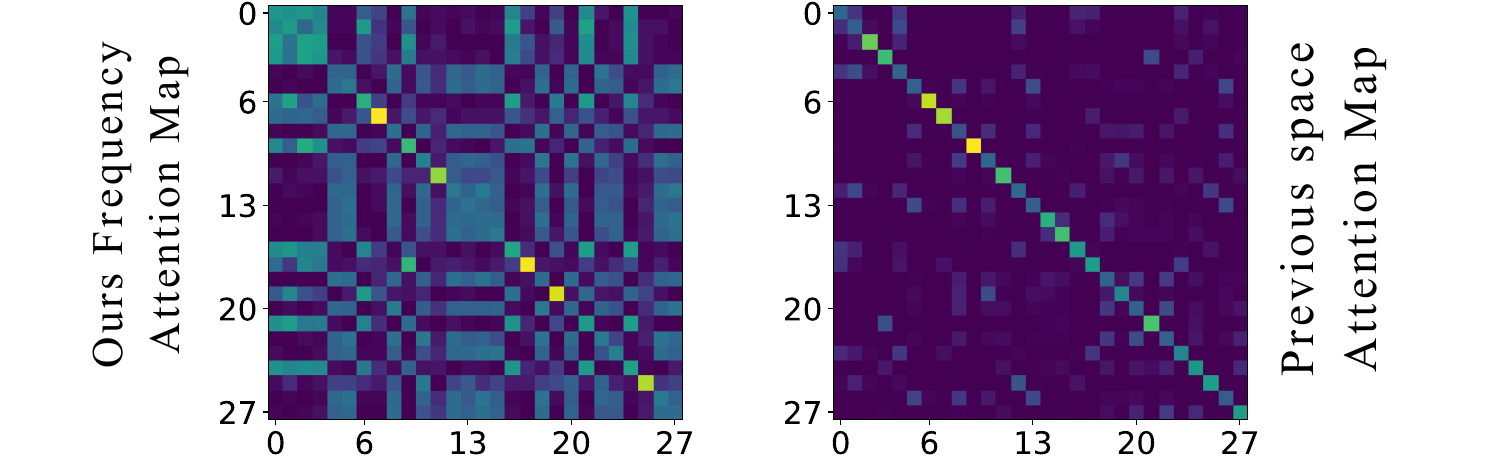}
   \end{center}
   \vspace{-12mm}
   \caption{\small Visualization of attention maps.}
   \vspace{-10mm}
   \label{fig:relation}
\end{wrapfigure} 

To further explore the relation between attention maps calculated in the method and the various correlation between spectral tokens, we visualize the heatmaps of space domain learning and the SAF, respectively. As depicted in Fig.~\ref{fig:relation}, the high correlation between frequency tokens in Fig.~\ref{fig:subfig1} brings the broad activated attention weights, while the low correlation between space tokens in Fig.~\ref{fig:subfig1} brings broad non-activated attention weights. To be concrete, highly correlated frequency tokens ensure that rich information from other relevant frequency tokens is comprehensively emerged. Conversely, poorly correlated space tokens lead to inefficient utilization of other space tokens and great interference by highly irrelevant tokens, thus harming the process of information extraction and evolution.

\vspace{-0.7em}
\section{Conclusion}
\label{sec:conclusion}
\vspace{-0.5em}

In this paper, we propose a valuable HFC prior derived from the statistical analyses of existent HSI datasets.
Leveraging the HFC, we formulate a frequency domain learning composed of a Spectral-wise self-Attention of Frequency and a Spectral-spatial Interaction of Frequency, which significantly enhances the HSI frequency exploration. By combining the frequency domain learning and the space domain learning, we architect the Correlation-driven Mixing Domains Transformer as deep prior for comprehensive prior exploitation. Plugging the delicate deep prior into prosperous unfolding architecture, we develop a correlation-driven mixing transformer unfolding framework for accurate HSI reconstruction. Experimental results reveal the superiority of our methods in image quality and computational efficiency over the SOTA methods.
Since the HFC widely exists in various HSIs, the HFC prior and the proposed method can be further leveraged in various HSI-based applications like spectral super-resolution, image classification, and biomedical analysis.


%
%
\bibliographystyle{splncs04}
\bibliography{main}
\end{document}